\documentclass[a4paper,11pt]{article}
\usepackage{pos}
\newcommand{\ab}{{\ensuremath\rm ab}}
\newcommand{\GeV}{{\ensuremath\rm GeV}}
\newcommand{\TeV}{{\ensuremath\rm TeV}}
\newcommand{\lam}{\lambda}
\newcommand{\lb}{\left(}
\newcommand{\rb}{\right)}
\newcommand{\fb}{{\ensuremath\rm fb}}
\newcommand{\eqn}{equation}

\title{News from Extended Scalar Sectors}
\author*[a]{Tania Robens}
\affiliation[a]{Rudjer Boskovic Institute,\\
  Bijenicka cesta 54, 10000 Zagreb, Croatia}

\emailAdd{trobens@irb.hr}
\abstract{In this proceeding contribution, I give a short overview on selected topics regarding extended scalar sector phenomenology. After a short overview on extended scalar sectors with light scalars at Higgs factories, I concentrate on the Inert Doublet model and recent studies exploring its discovery potential at lepton colliders. \\ RBI-ThPhys-2026-08}

\FullConference{Proceedings of the Corfu Summer Institute 2025 "School and Workshops on Elementary Particle Physics and Gravity" (CORFU2025)\\
27 April - 28 September, 2025\\
Corfu, Greece\\}

\begin{document}
\maketitle

\section{Introduction}
In this short proceedings contribution, I briefly comment on new results that have been obtained for models with extended scalar sectors, with a focus on lepton colliders. After the discovery of the Higgs boson, one important question is whether the electroweak sector realized in nature is the one predicted by the SM, or whether an extended scalar sector is realized in nature. The latter can have various additional advantages of e.g. solving puzzles of the SM, such as the existance of dark matter, the evolution of the universe, or the stability of the electroweak vacuum. A thorough discussion of this is beyond the scope of this proceeding contribution and we refer the reader to e.g. \cite{Robens:2025nev} for further discussions.

After the HL-LHC, the next collider that should be realized in nature is a so-called Higgs factory, that allows to probe the properties of the SM scalar boson up to high precision. In addition, of course, it is interesting to investigate which kind of new physics scenarios can be investigated at such colliders. If we assume center-of-mass energies of around $250\,\GeV$ for such machines, naturally additional particles need to be light in order to be produced onshell in reasonable rates.

In the following, I will therefore first discuss general searches as well as some recent studies of light scalars at Higgs factories. I will then turn to a specific model, the Inert Doublet Model (IDM), and present two recent studies for this at a low energy $e^+\,e^-$ collider as well as a \TeV scale muon collider. I then finish with some brief conclusion.

\section{Light scalars at lepton colliders}

\begin{center}
  \begin{figure}[htb!]
    \begin{center}
    \includegraphics[width=0.5\textwidth]{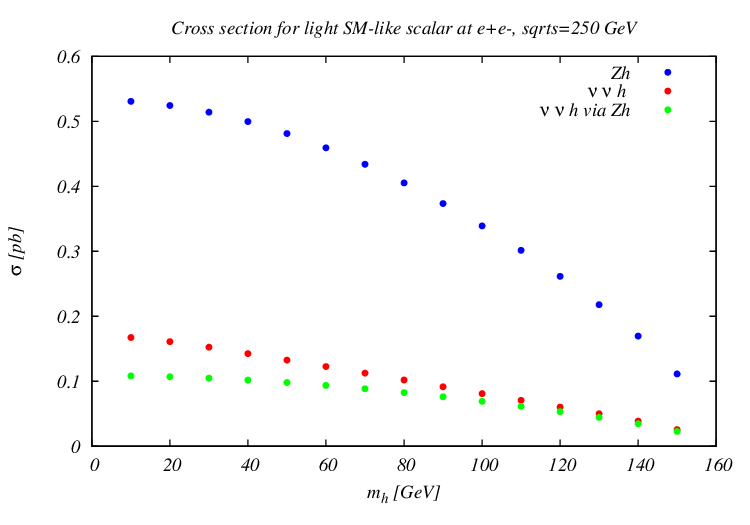}
\caption{\label{fig:prod250} { Leading-order production cross sections for $e^+\,e^-\,\rightarrow\,Z\,h$ {\sl (blue)} and $e^+\,e^-\,\rightarrow\,h\,\nu_\ell\,\bar{\nu}_\ell$ {\sl (red)} production at an $e^+\,e^-$ collider with a com energy of 250 \GeV~using Madgraph5 for an SM-like scalar $h$. The contribution of $Z\,h$ to $\nu_\ell\,\bar{\nu}_\ell\,h$ using a factorized approach for the Z decay is also displayed {\sl (green)}. Taken from \cite{Robens:2024wbw}.}}
    \end{center}
    \end{figure}
\end{center}

At the center-of-mass (com) energies of Higgs factories, Higgs strahlung is the dominant production mode for single scalar production \cite{Abramowicz:2016zbo}. In figure \ref{fig:prod250}, we show leading-order predictions for $Zh$ production at $e^+e^-$ colliders for low mass scalars which are Standard Model (SM)-like, using Madgraph5 \cite{Alwall:2011uj}, for a center-of-mass energy of 250 \GeV. The $e^+e^-\,\rightarrow\,h\,\nu_\ell\,\bar{\nu}_\ell$ process contains contributions for both scalar strahlung and VBF type topologies. In order to disentangle these, we also display the expected rates from the former for this final state using a factorized approach. For higher scalar masses the dominant contribution stems from $Z\,h$ production.

We now turn to specific final states and novel results for the investigation of their discovery potential. In \cite{deBlas:2024bmz}, some target processes were identified that should be reanalysed for the European Strategy input. This lead to several new studies concentrating mainly on scalar-strahlung processes with $b\bar{b}$ and $\tau^+\tau^-$ final states. A comparative study of several of these searches is also shown.

\subsection*{$b\,\bar{b}$ final states}

Light scalars that are SM-like typically predominantly decay into $b\,\bar{b}$ final states. This channel has already been explored in detail at the LEP experiments (see \cite{OPAL:2002ifx,ALEPH:2006tnd} for a concise summary of results). In these searches, two different options of treating the decay modes of the additional scalars exist: one just uses the general recoil against the $Z$ and does not take specific decays of the scalar into account. The second option also makes use of the final states stemming from the additional scalar\footnote{For a comparison using a LEP recast see e.g. \cite{Drechsel:2018mgd}.}. We here focus on the latter using $b\bar{b}$ final states.

\begin{center}
  \begin{figure}[htb!]
    \begin{center}
      \includegraphics[width=0.6\textwidth]{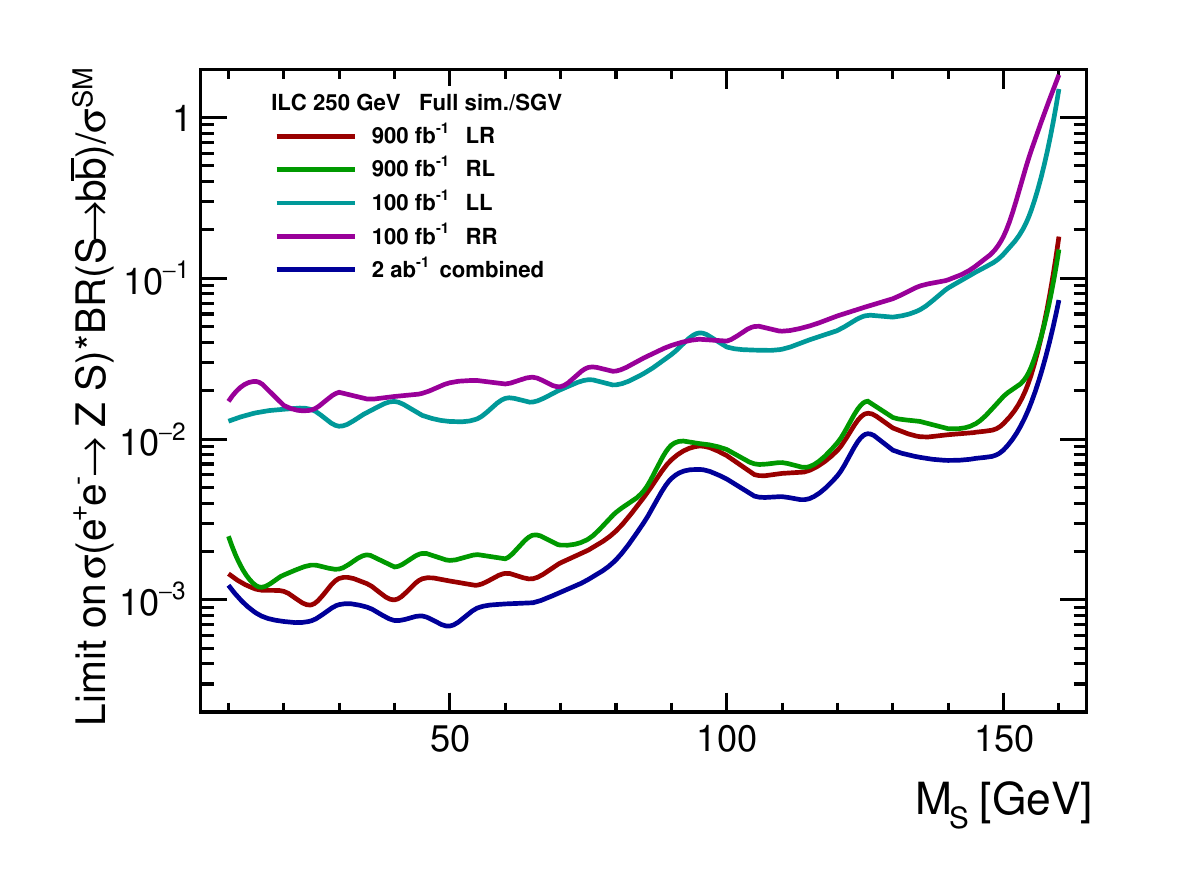}
    \end{center}
    \caption{\label{fig:bb} Taken from \cite{Altmann:2025feg} for $b\bar{b}$ final states, for various initial state polarization combinations at various integrated luminosities as well as a combination at $2\,\ab^{-1}$. Results are given for production times decay, normalized to the SM production cross section at that mass.}
    \end{figure}
\end{center}

In figure \ref{fig:bb}, we display the results of a detailed comparison at a 250 \GeV~ ILC, including effects of introducing initial state polarization configurations. The best results correspond to a combination at an integrated luminosity of $2\,\ab^{-1}$. The constrained quantity can be translated to the squared rescaling for $Z\,S$ production times $S\,\rightarrow\,b\,\bar{b}$ branching ratio in many new physics scenarios.

\subsection*{$\tau^+\tau^-$ final states}

\begin{center}
  \begin{figure}[htb!]
    \begin{center}
      \includegraphics[width=0.49\textwidth]{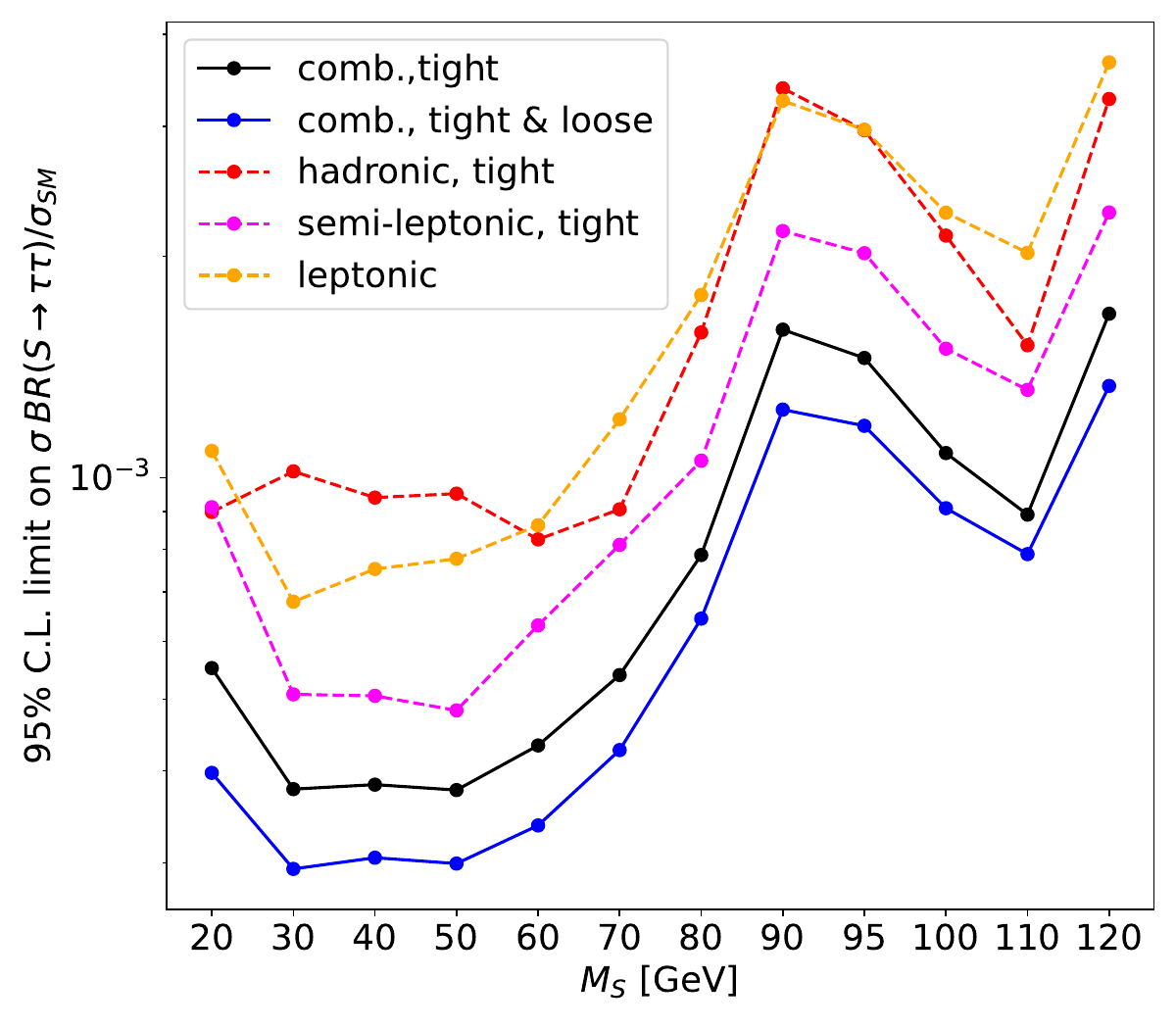}
    \end{center}
    \caption{\label{fig:tautau} Taken from \cite{Altmann:2025feg} for $\tau^+\,\tau^-$ final states, for an ILC at 250 \GeV~ and an integrated luminosity of $2\,\ab^{-1}$. Result are shown for various final states and selection criteria for the di-tau decay products. See \cite{Altmann:2025feg} for details. }
    \end{figure}
  \end{center}

Another focus promoted in \cite{deBlas:2024bmz} are processes where the additional scalar decays into di-tau final states. The results of such a study are shown in figure \ref{fig:tautau}, where various decay and selection modes of the scalar decay products are taken into account, at a center of mass energy of 250 \GeV~ and an ILC setup with an integrated luminosity of $2\,\ab^{-1}$. Best results can be obtained in an overall combination of all selection criteria searches.

\begin{center}
  \begin{figure}[htb!]
    \begin{center}
      \includegraphics[width=0.6\textwidth]{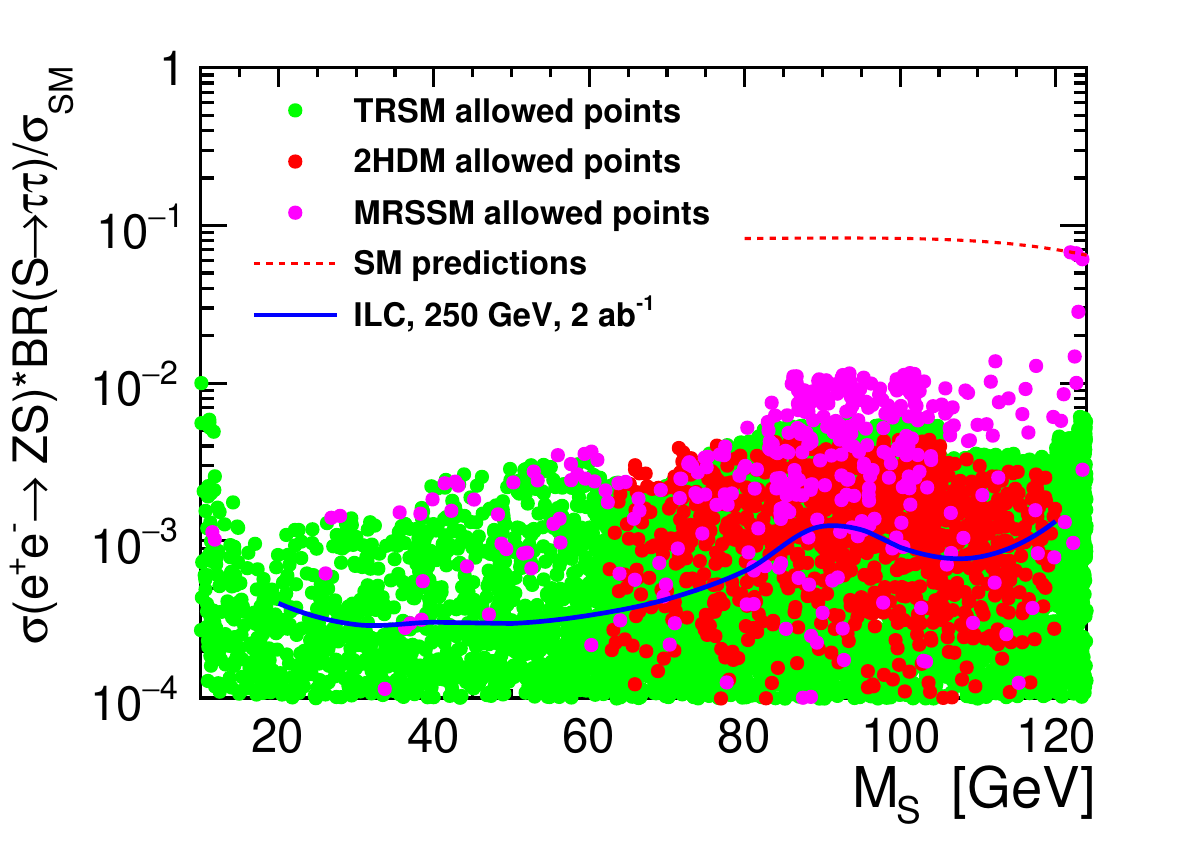}
    \end{center}
    \caption{\label{fig:taucomp} Taken from \cite{Altmann:2025feg} for $\tau^+\,\tau^-$ final states, viability of different models.}
    \end{figure}
\end{center}

In this context, one can also ask which models can actually still be constrained by such searches. 
Therefore,  in figure \ref{fig:taucomp} we show the possible reach of an additional scalar decaying into ditau final states and compare to the parameter space recently still allowed by theoretical and experimental constraints for various new physics scenarios:  the TRSM, a model where the SM scalar sector is enhanced by two real singlets \cite{Robens:2019kga,Robens:2022nnw}, a 2HDM, with the scan obtained using thdmTools \cite{Biekotter:2023eil}, as well as the MRSSM which corresponds to the MSSM with a continuous R-symmetry (see e.g. \cite{Diessner:2019bwv,Kalinowski:2024uxe}). It is obvious that given currently allowed regions, this search can severly constrain the parameter space of the models.

\begin{center}
  \begin{figure}[htb!]
    \begin{center}
      \includegraphics[width=0.8\textwidth]{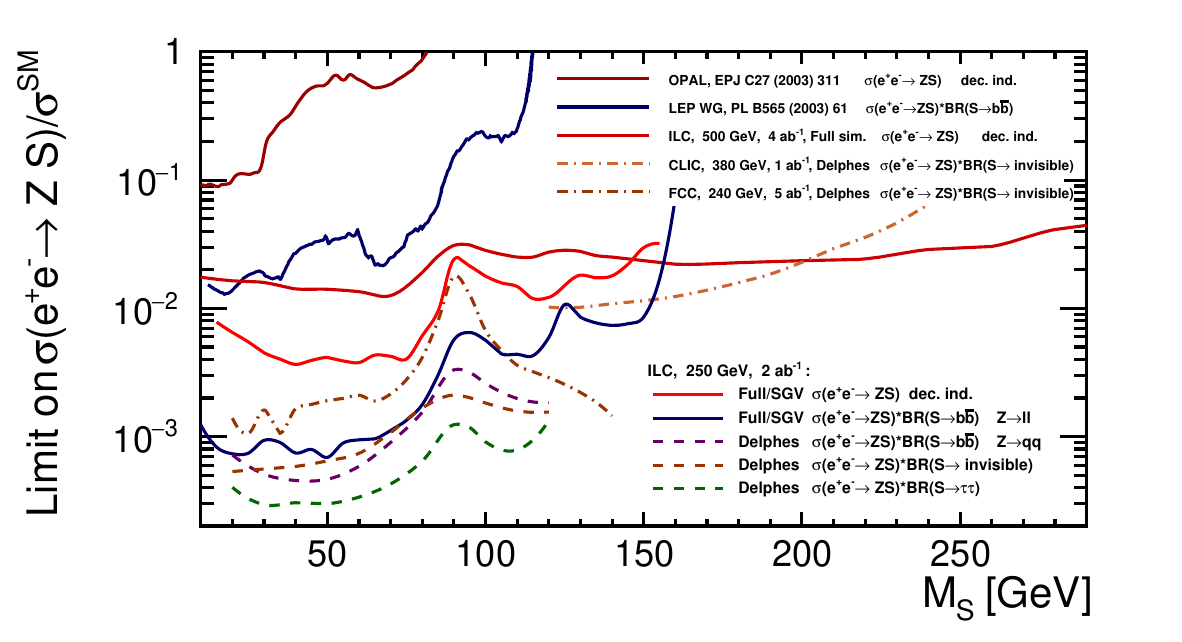}
    \end{center}
    \caption{\label{fig:combi} Taken from \cite{Altmann:2025feg} for several final states, reach of several searches using a variety of decay modes, center-of-mass energies, and detector simulation approaches. Shown are also the experimental bounds achieved by the LEP collaborations \cite{OPAL:2002ifx,ALEPH:2006tnd}.}
    \end{figure}
  \end{center}

Several other final states, in particular also searches with invisible decays of the additional scalar, have been documented in \cite{Altmann:2025feg}. In figure \ref{fig:combi}, we show the comparison of various of such searches and their reach for the normalized rates. Note that some results are given for a rescaled production cross section only, while other include a factorized decay for the additional scalar. Higher center-of-mass energies naturally lead to higher mass reaches. For all final states, the 250 \GeV~ ILC studies at 2 $\ab^{-1}$ supersede previous studies for the corresponding mass range. In general the di-tau final states appears to be most sensitive in the studies considered here. Also displayed are actual search results from LEP.

\section{Inert Doublet Model}

The Inert Doublet Model \cite{Deshpande:1977rw,Barbieri:2006dq,Cao:2007rm} is a two Higgs doublet model with an exact $\mathbb{Z}_2$ symmetry that renders the lightest scalar of the second doublet stable and therefore provides a dark matter candidate. The so-called dark scalars typically decay via electroweak gauge bosons to final states including the dark matter candidate. The potential is given by

\begin{align}
    V_\text{IDM} =&\ \mu_1^2 |\Phi_1|^2 + \mu_2^2 |\Phi_2|^2 + \frac{1}{2} \lambda_1 |\Phi_1|^4 + \frac{1}{2} \lambda_2 |\Phi_2|^4 + \lambda_3 |\Phi_1|^2 |\Phi_2|^2 + \lambda_4 |\Phi_1^\dagger \Phi_2|^2 \nonumber\\
    &+ \frac{1}{2} \lambda_5 \left[(\Phi_1^\dagger\Phi_2)^2+\text{h.c.}\right], 
\end{align}
with
\begin{align}
    \Phi_1=\begin{pmatrix}
      G^+\\
      \frac{1}{\sqrt{2}}(v+h+iG^0)
    \end{pmatrix}\,,
    \quad \text{and} \quad
    \Phi_2=\begin{pmatrix}
      H^+\\
      \frac{1}{\sqrt{2}}(H+iA)
    \end{pmatrix}\,,
\end{align}
and where $h\,H,\,A,\,H^+$ denote the physical mass eigenstates, and all couplings are taken to be real.

The results presented here are based in the scan developed in the context of \cite{Ilnicka:2015jba}  and subsequently updated in \cite{Kalinowski:2018ylg,Dercks:2018wch,Kalinowski:2020rmb,Braathen:2024ckk,Bal:2025nbu,Lahiri:2025opz}. We refer the reader to these works for more details.

Due to the exact $\mathbb{Z}_2$ symmetry, electroweak symmetry breaking originates purely from the SM like doublet and proceeds as in the SM. The model features seven free parameters after electroweak symmetry breaking which we choose to be
\begin{\eqn*}
  M_h,\,M_H,\,M_A,\,M_{H^\pm},\,v,\,\lam_2,\,\lam_{345}\,\equiv\,\lam_3+\lam_4+\lam_5.
  \end{\eqn*}

Two parameters, namely $M_h\,\sim\,125\,\GeV$ and $v\,\sim\,246\,\GeV$, are fixed by current experimental findings.
The lightest new scalar features as the DM candidate. In this work, we chose $H$ to play this role. It is also important to note that $\lam_{345}$ determines the $h\,H\,H$ coupling and therefore is an important parameter for dark matter related constraints.

\begin{center}
  \begin{figure}[htb!]
    \begin{center}
      \includegraphics[width=0.45\textwidth]{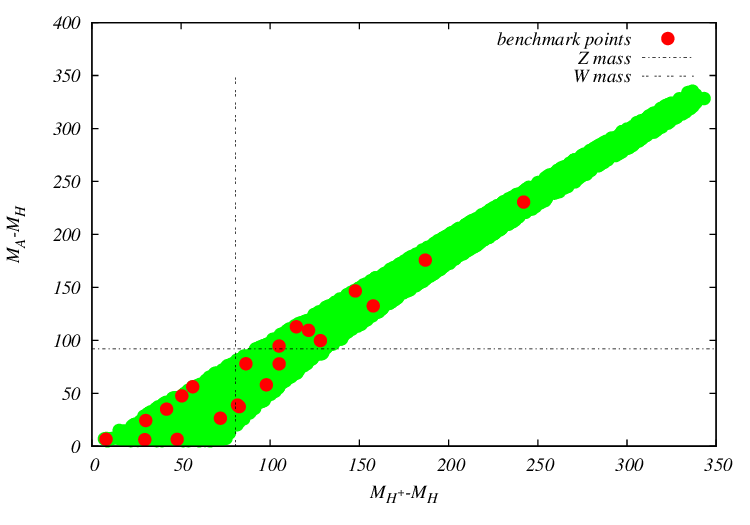}
       \includegraphics[width=0.45\textwidth]{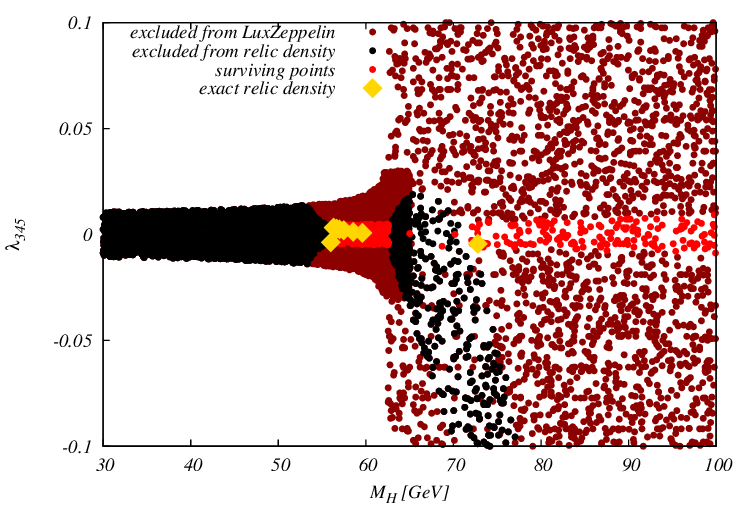}
    \end{center}
    \caption{\label{fig:idmex} {\sl (Left):} Available parameter space in the $\lb M_{H^\pm}-M_H,\;M_A-M_H  \rb$ parameter plane after all constraints are taken into account. The unstable new scalars are relatively mass degenerate, mainly due to constraints on couplings in the potential as well as bounds from electroweak precision constraints. Figure taken from \cite{Kalinowski:2020rmb}. Also shown are benchmark points proposed in this work. {\sl (Right):} Allowed parameter space in the $\lb M_H,\;\lam_{345} \rb$ parameter plane. For $M_H\,\leq\,62.5\,\GeV$, the contour stems from applying constraints from the $h\,\rightarrow\,\text{invisible}$ branching ratio. We see that in combination with constraints from relic density this leads to a lower mass scale for DM. Figure taken from \cite{Robens:2022jrs}, as an update of a figure presented in \cite{Ilnicka:2018def}. Dark matter constraints are taken from \cite{Planck:2018vyg,LZ:2024zvo}. }
    \end{figure}
\end{center}

The model is subject to a large number of theoretical and experimental constraints that are discussed in the above work and will not be repeated here. However, we briefly want to focus on scenarios where indeed the constraints lead to generic distinct features in parameter space. Two examples for this are shown in figure \ref{fig:idmex}.

In general it is not easy to find two-dimensional parameter planes where constraints render clear contours. Examples given here include a combination of bounds on couplings and electroweak precision constraints that lead to a relatively narrow region in the unstable mass parameter plane, as well as clear constraints on the coupling $\lam_{345}$ as a function of the dark matter mass.

\begin{center}
  \begin{figure}[htb!]
    \begin{center}
      \includegraphics[width=0.45\textwidth]{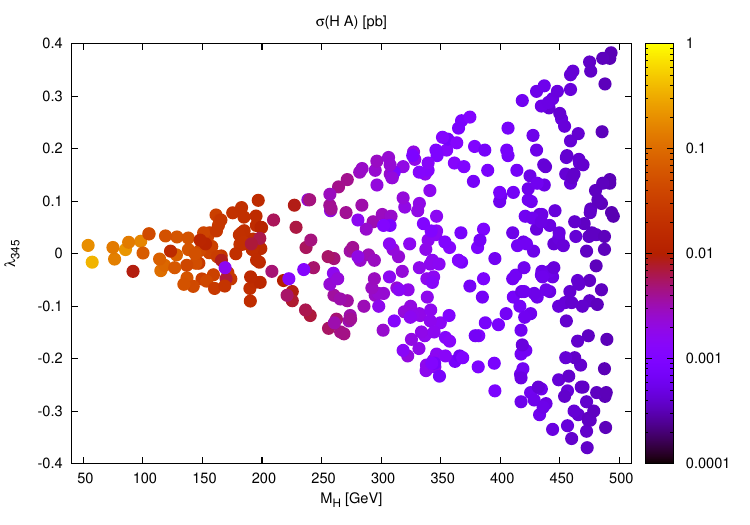}
      \includegraphics[width=0.45\textwidth]{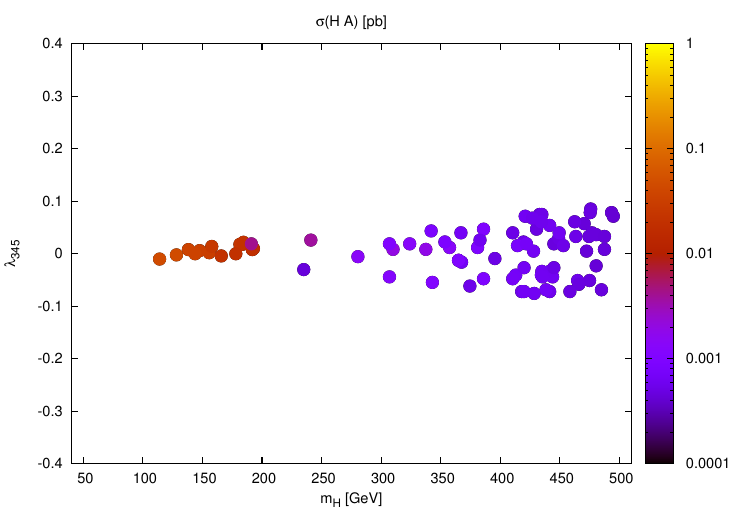}
       \includegraphics[width=0.45\textwidth]{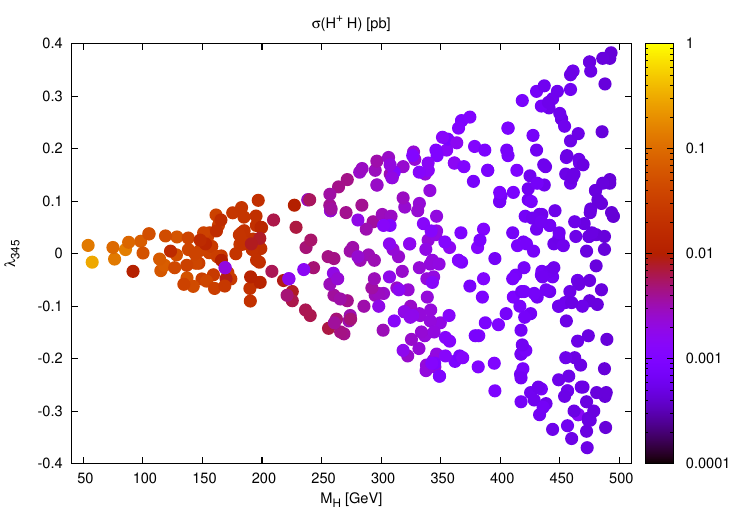}
      \includegraphics[width=0.45\textwidth]{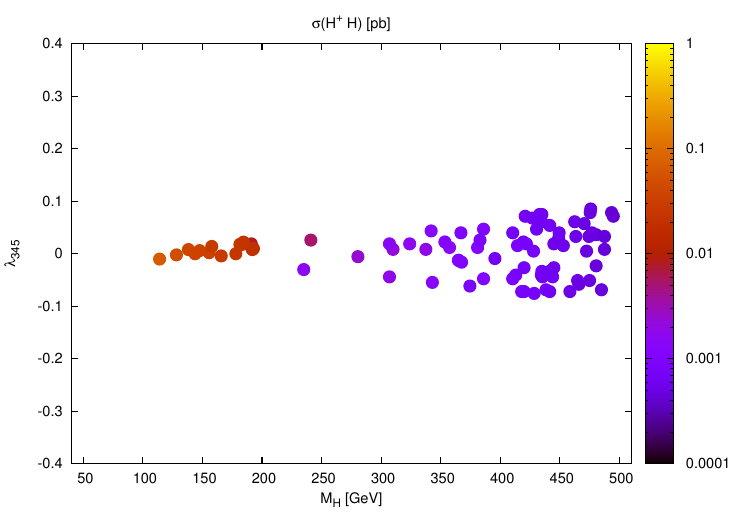}
    \end{center}
    \caption{\label{fig:dirdet}. Constraints in the $\lb M_H,\;\lam_{345} \rb$ plane  using direct detection results from LUX and LUX-ZEPLIN, respectively. Color coding corresponds to production cross sections for $H\,A$ {\sl (top) }and $H^\pm\,H$ pair-production at a 13 \TeV~ LHC and are not of importance here. Taken from \cite{Ilnicka:2015jba} in the original version for LUX results.}
    \end{figure}
\end{center}

Another important example on the importance of experimental bounds is the application of dark matter constraints from direct detection on that same plane. For this, we compare the available parameter space that was available using LUX data \cite{LUX:2013afz} with bounds from the most recent measurements from LUX-ZEPLIN \cite{LZ:2024zvo}. The results of this comparison are displayed in figure \ref{fig:dirdet}.

It becomes obvious that the direct detection bounds pose important constraints on the models parameter space.

\subsection{At a Higgs factory}
\begin{center}
  \begin{figure}[htb!]
    \begin{center}
     \includegraphics[width=0.3\textwidth]{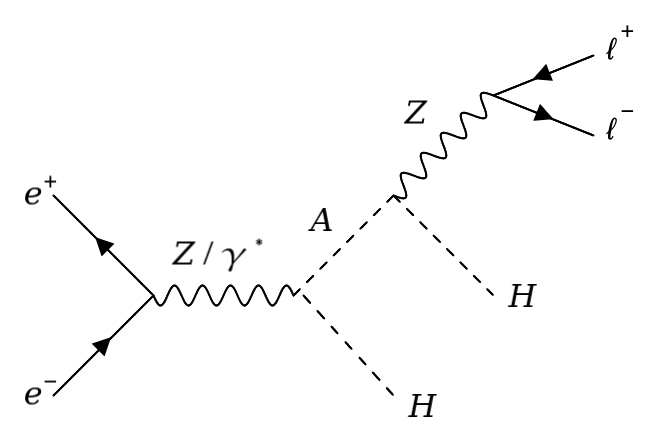}\hfill
        \includegraphics[width=0.3\textwidth]{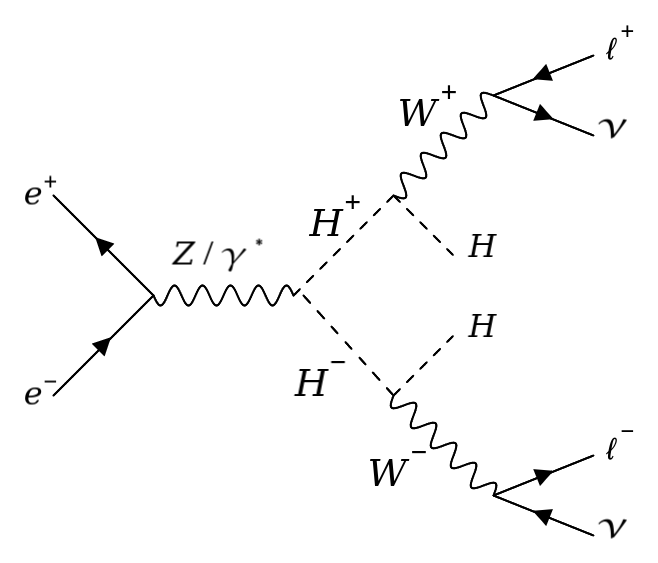}\hfill
        \includegraphics[width=0.3\textwidth]{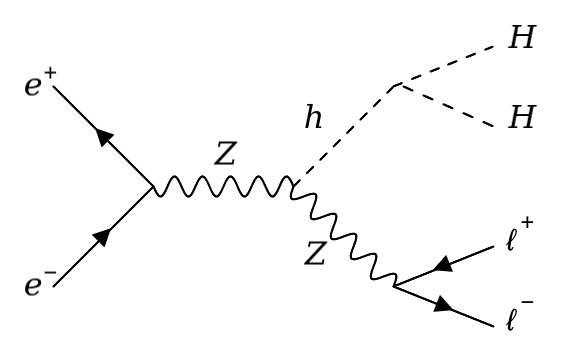}
    \end{center}
    \caption{\label{fig:feyn} Dominant production mechanisms for dilepton and missing energy final states at a lepton collider within the Inert Doublet Model. The original target signature is given by the first diagram, while the second and third correspond to additional contributions to the same experimental signature. These can in principle be suppressed by choosing the free parameters of the model accordingly.}
    \end{figure}
\end{center}

We now briefly discuss results from the investigation of the Inert Doublet Model at Higgs factories, that are available in \cite{Bal:2025nbu}. That work concentrates on final states with opposite sign same flavour leptons and missing energy. Corresponding dominant Feynman diagrams for the production are given in figure \ref{fig:feyn}.

The target signature here is given by $A\,H$ production and subsequent decays. It is interesting to see that the additional diagrams in figure \ref{fig:feyn} can in principle be suppressed using adequate choices of the free parameters of the model. The additional contribution from charged scalar production e.g. becomes smaller for larger masses $M_{H^\pm}$; on the other hand, a larger mass gap to the decay products typically leads to a larger branching ratio into that final state. The third diagram, on the other hand, is directly proportional to $\lam_{345}$ and can therefore be suppressed for small values of that coupling.

In the above reference, three scenarios are defined in order to exemplify the importance of the separate contributions:

\begin{eqnarray*}
(S1)&&M_{H^\pm}\,=\,M_A,\,\lam_{345}\,=\,0;\\
(S2)&&M_{H^\pm}\,=\,M_A,\,\lam_{345}\,=\,\lam_{345}^\text{max};\\
(S3)&&M_{H^\pm}\,=\,M_{H^\pm}^\text{max},\,\lam_{345}\,=\,\lam_{345}^\text{max}
\end{eqnarray*}

Comparisons between the first and second scenario allow for an estimate of the contributions from Higgs strahlung, while the third scenarios investigates the contributions from charged scalar production. Maximal allowed values for $\lam_{345}$ and $M_{H^\pm}$ depended on the bounds that were applied and can be found in the original publication.

\begin{center}
  \begin{figure}[htb!]
    \begin{center}
\includegraphics[width=0.45\textwidth]{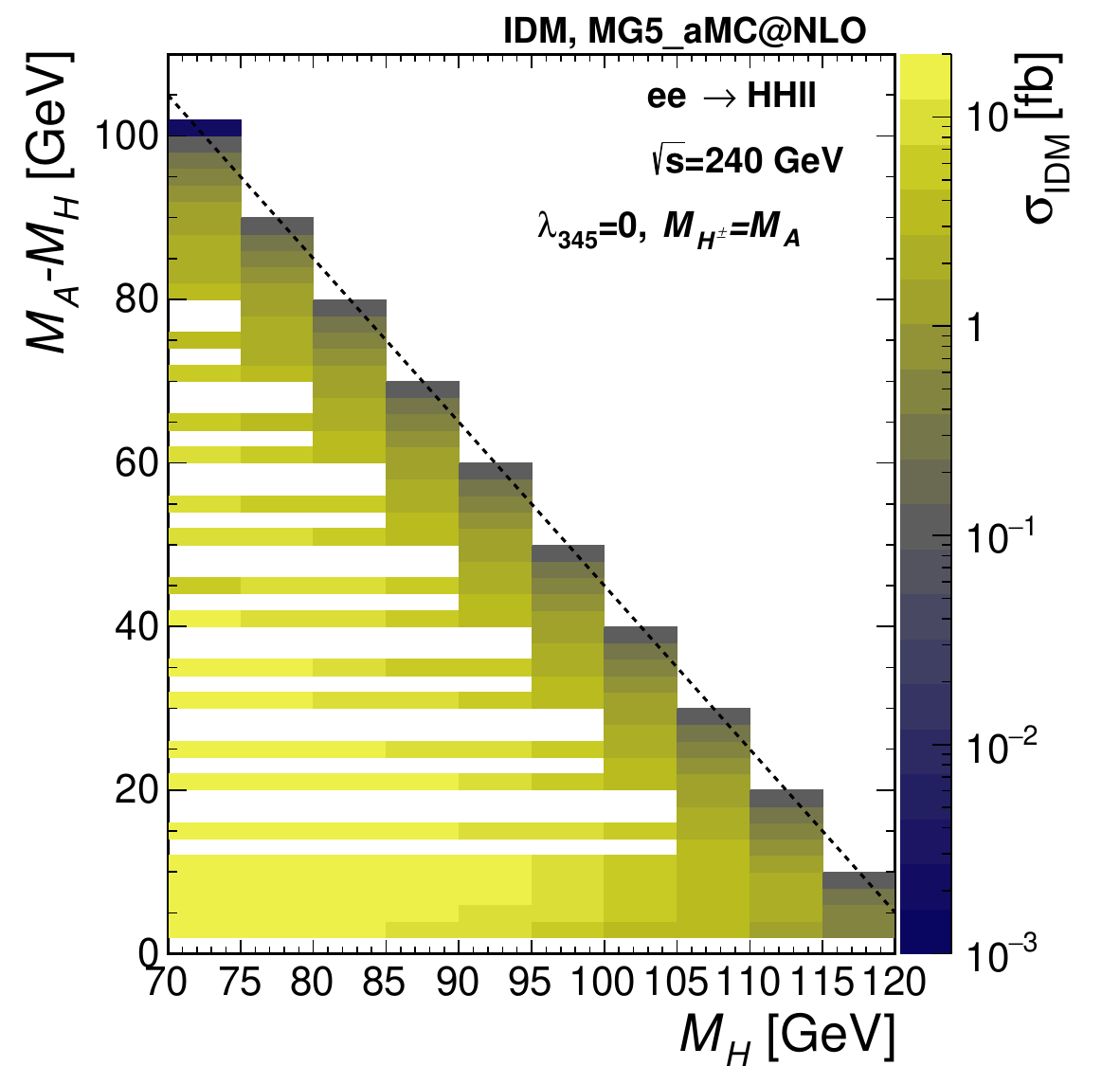}\hfill
\includegraphics[width=0.45\textwidth]{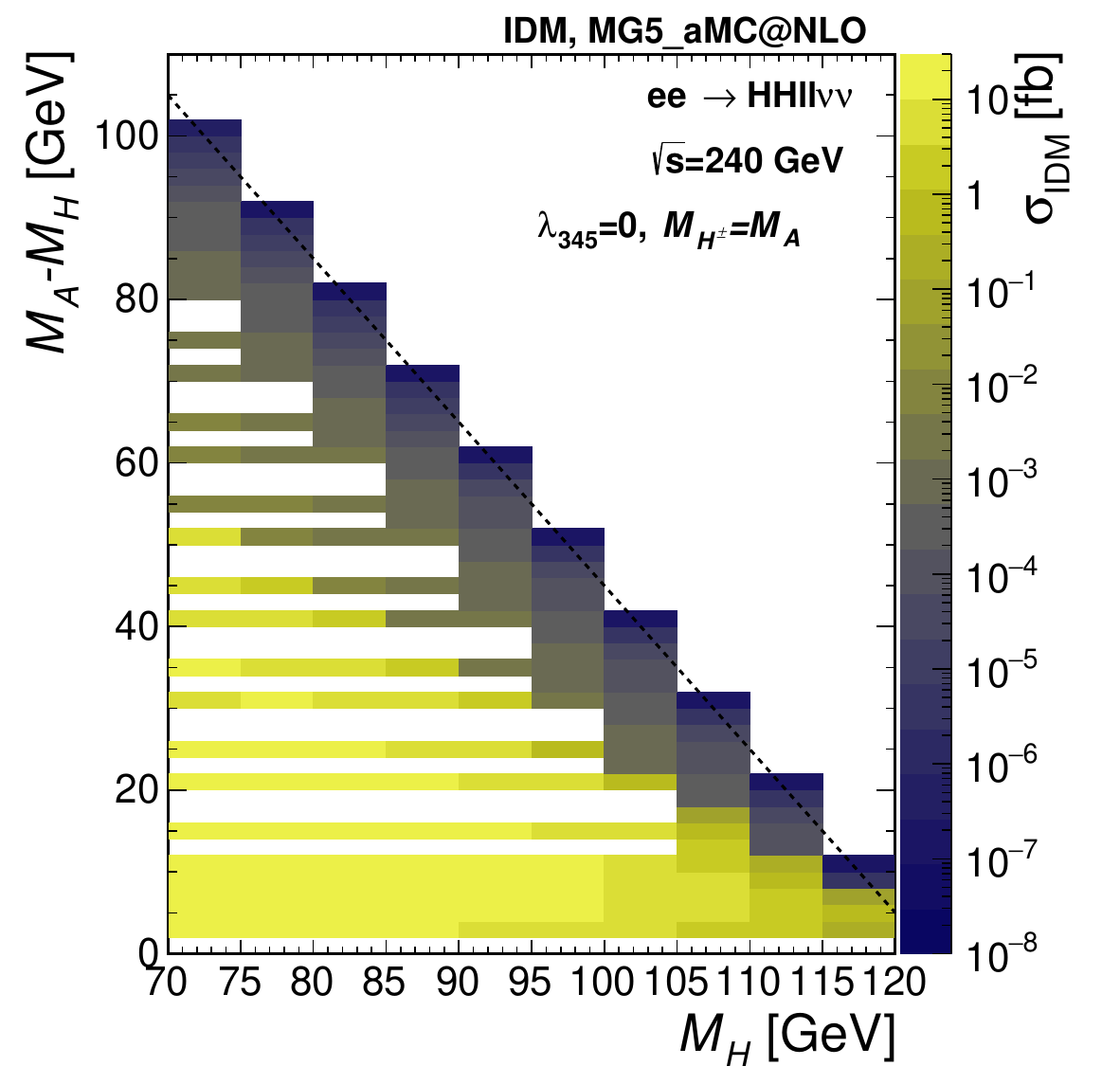}
    \end{center}
    \caption{\label{fig:xsecs} Cross section from the MG5 aMC@NLO simulation at 240 \GeV, in the $\lb M_H;\;M_A-M_H \rb$ plane, in the first scenario. Cross sections are given in \fb. {\sl (Left):} dilepton and $HH$ final states; {\sl (right)} including additional neutrinos. The dashed line shows
the kinematic limit for on-shell production. Taken from \cite{Bal:2025nbu}.}
    \end{figure}
\end{center}

In figure \ref{fig:xsecs}, we present the cross section values for the respective contributions at a lepton collider with a center of mass energy of 240 \GeV, with an inital $p_\perp$ cut of 0.5 \GeV~ for the leptons in the first scenario. As expected, the cross sections are mainly dominated by kinematics determined by the mass scales.

\begin{center}
  \begin{figure}[htb!]
    \begin{center}
 \includegraphics[width=0.4\textwidth]{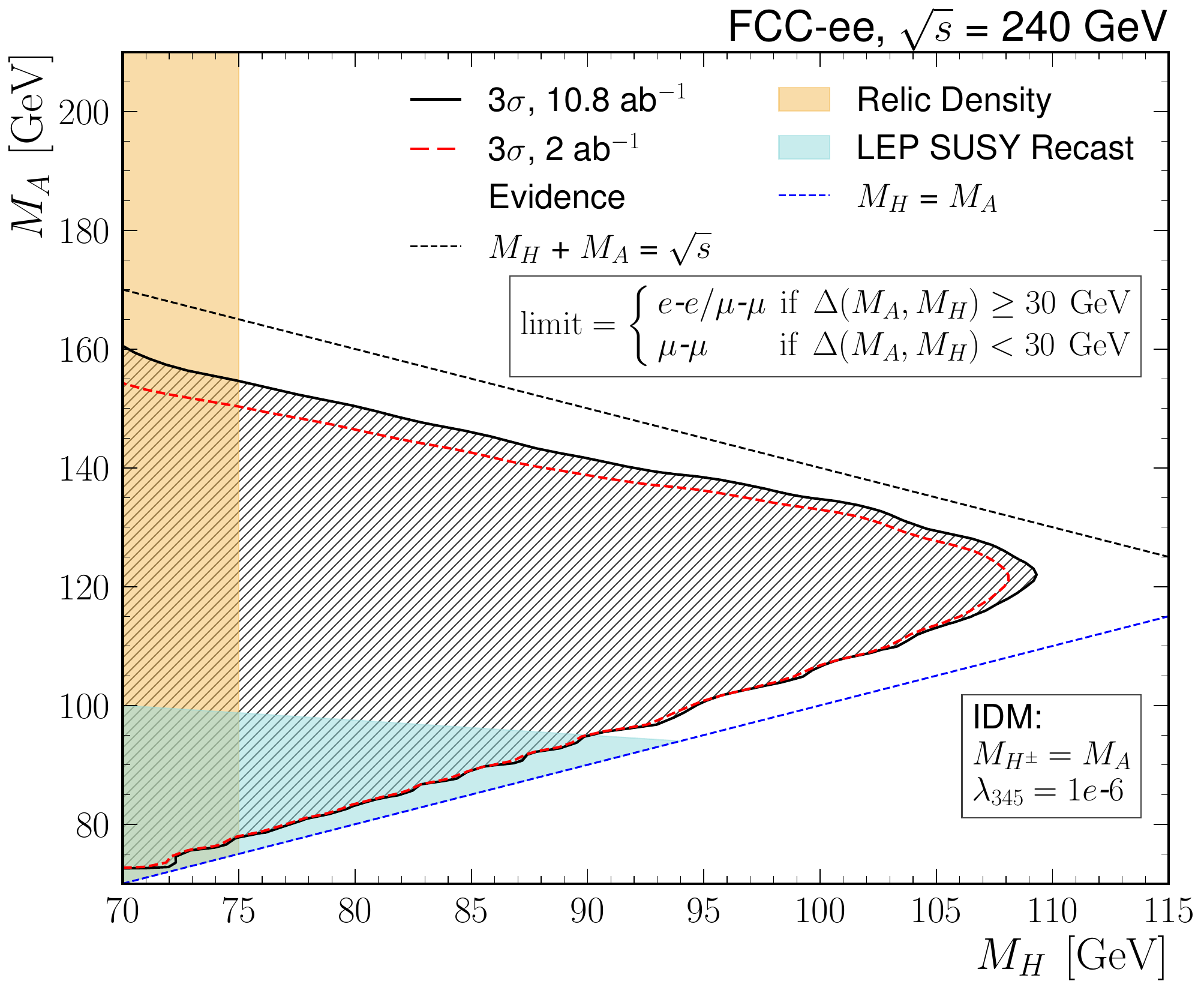} \hspace{3mm}    
    \includegraphics[width=0.4\textwidth]{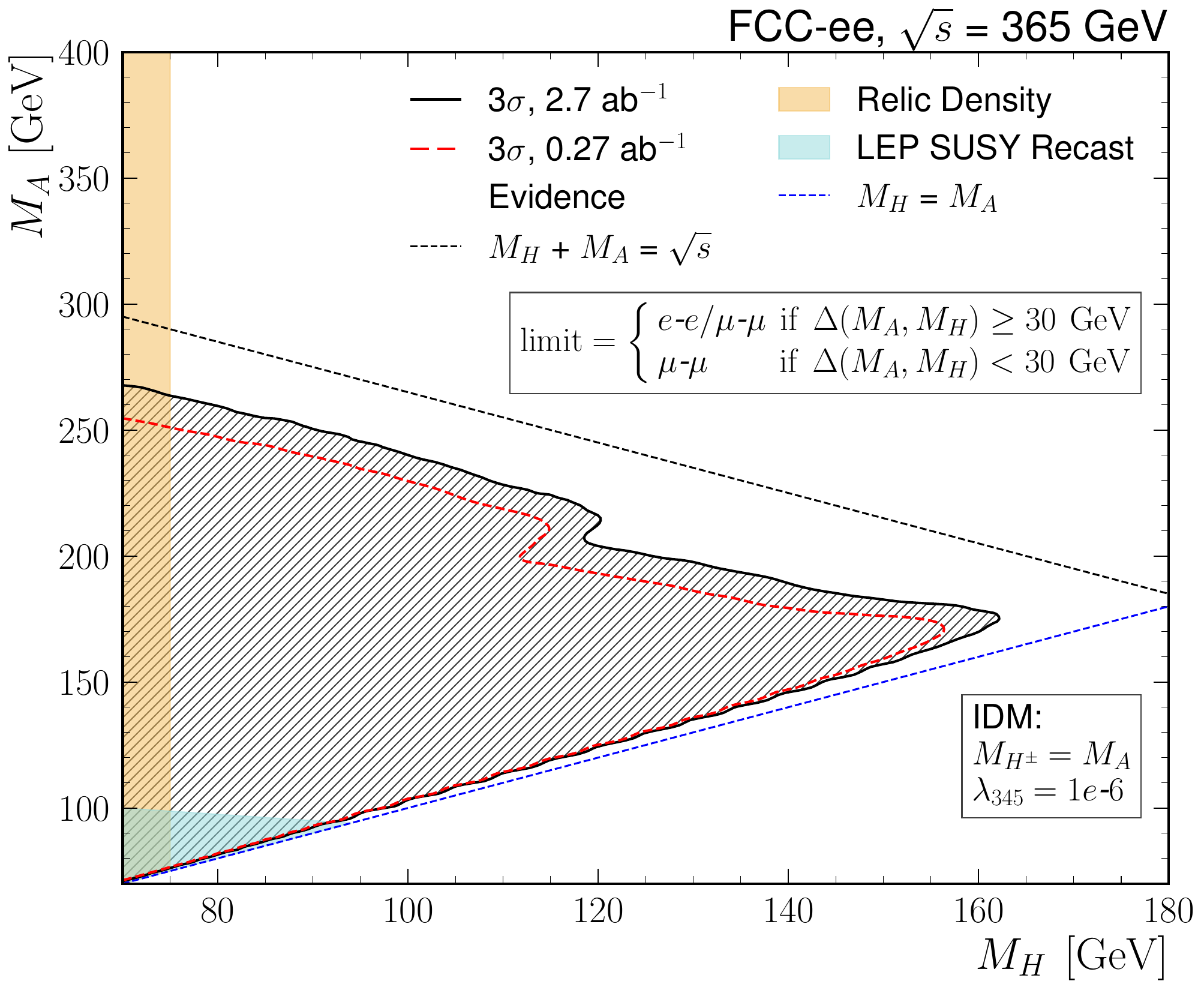}\\
  \includegraphics[width=0.4\textwidth]{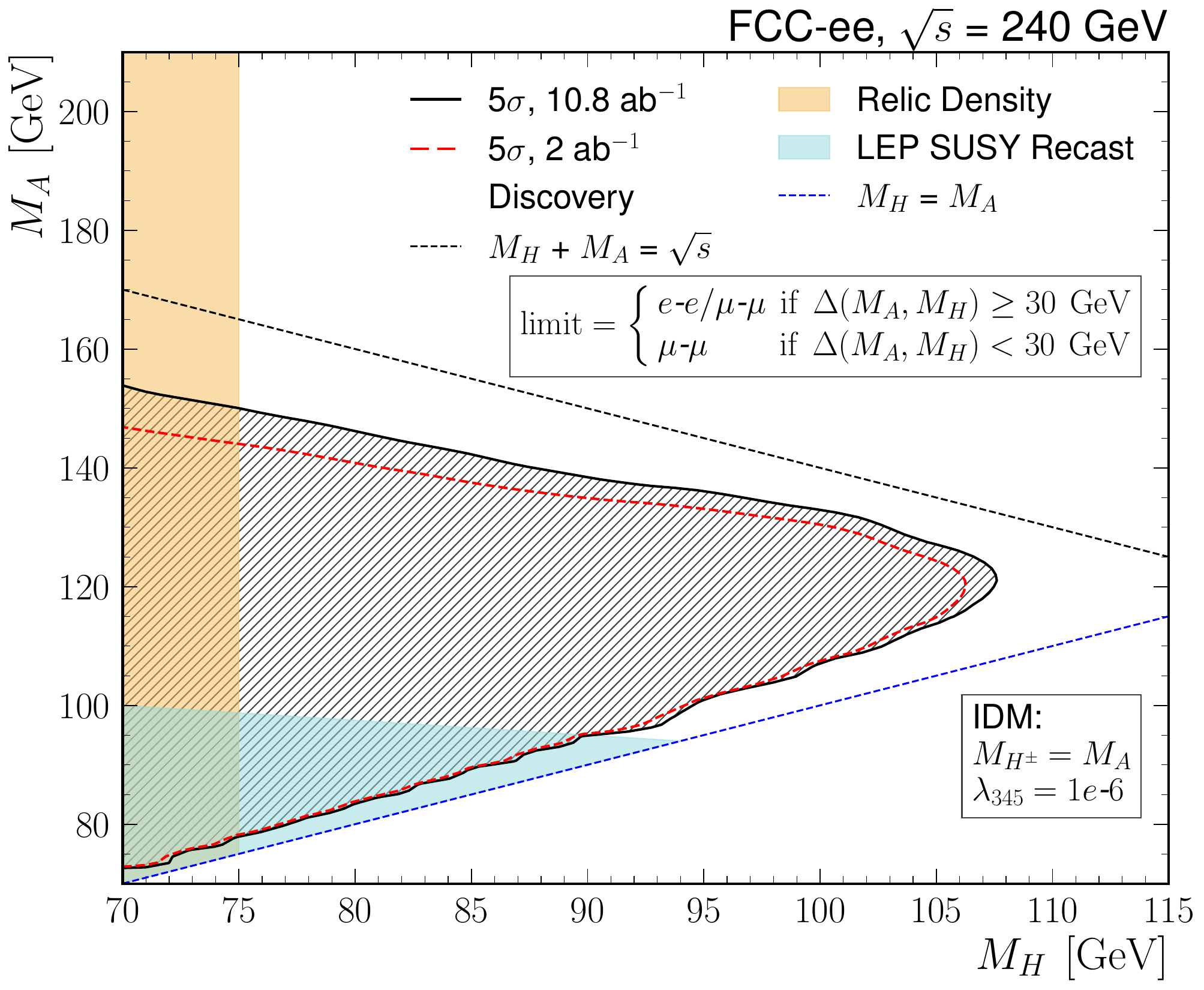} \hspace{3mm}
    \includegraphics[width=0.4\textwidth]{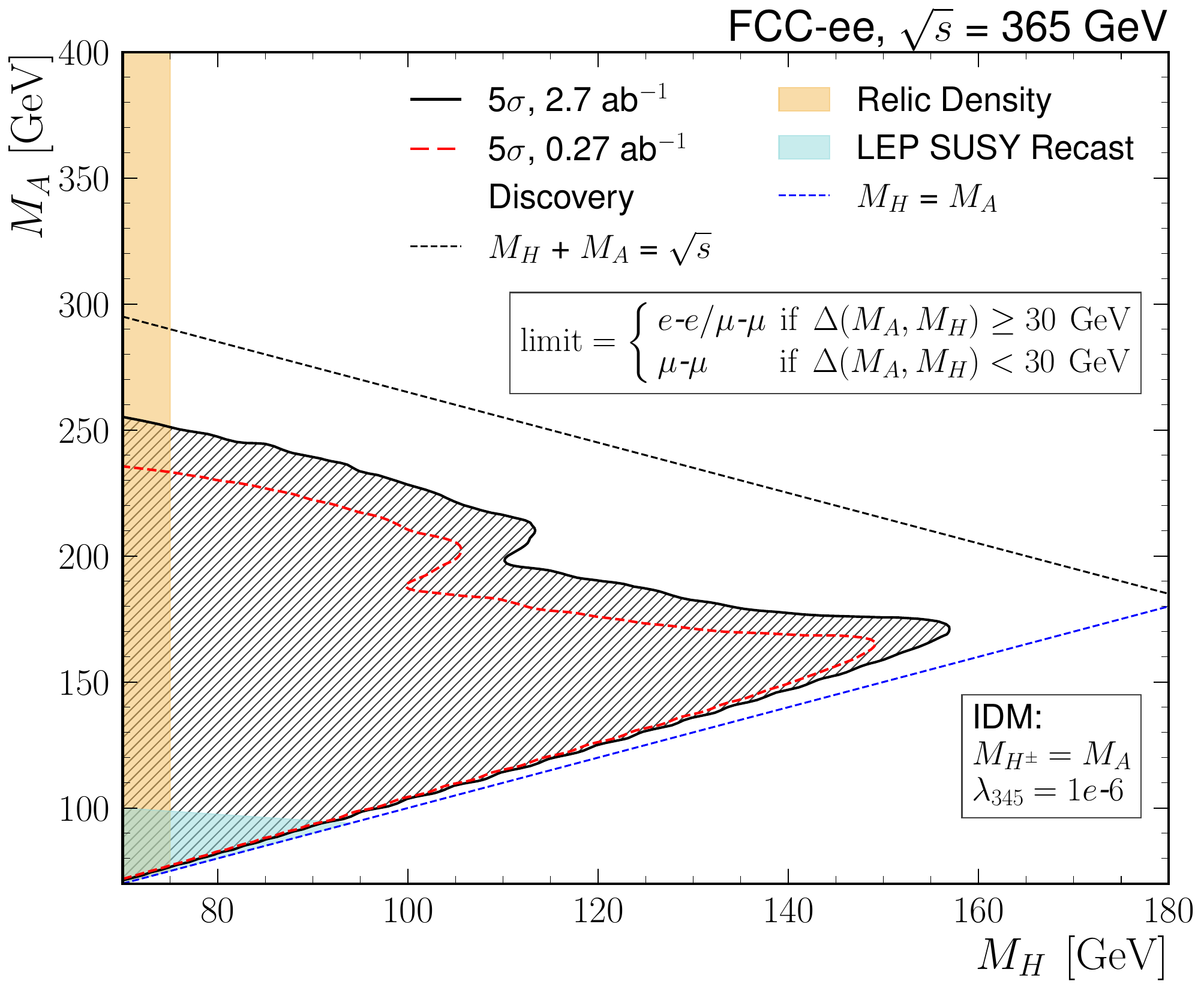}
    \end{center}
    \caption{\label{fig:ress} Evidence {\sl (upper row)} and discovery potential {\sl (lower row)} for dilepton and missing energy final states in the IDM, for S1 as defined in the text, for center of mass energies of 240 \GeV~ and 365 \GeV~ and various integrated luminosities, respectively.}
    \end{figure}
\end{center}

The full analysis includes a dedicated simulation of backgrounds, detector effects, as well as a set of preselection cuts and the training of a parametric neural network. We refer the reader to the original publication for details. Instead, we here chose to only show the final results; these are given in figure \ref{fig:ress}.
In addition to the expected evidence an discovery potential of an $e^+e^-$ collider at different center of mass energies, the figure displays kinematic bounds as well as constraints from dark matter and LEP searches \cite{Lundstrom:2008ai}. We see that the sensitivity for evidence nearly covers the whole parameter space. Regarding discovery, larger values of $M_A$ are disfavoured.

\subsection{At a muon collider}
We now turn to the discovery potential of a high-energy muon collider with a center of mass energy of 10 \TeV. This subsection summarizes the work presented in \cite{Braathen:2024ckk}.

The advantage of a high energy lepton collider is that for \TeV~ scale center of mass energies they basically turn into a vector boson collider, with VBF type topologies giving the largest rates. This makes processes accessible that are e.g. proportional to Yukawa couplings in the normal production modes for s-channel contributions. In the IDM, the $A\,A$ final state can only be produced via an s-channel 125 \GeV scalar or quartic couplings. In order to enhance the former, it is instructive to consider the VBF-type production of these particles at high energy lepton colliders.

Another important quantity for this process is the $h_{125}\,A\,A$ coupling determined by

\begin{\eqn}\label{eq:lam345b}
  \bar{\lam}_{345}\,\equiv\,\lam_{345}\,+\,2\,\frac{M_A^2-M_H^2}{v^2}.
\end{\eqn}
The second term enables relatively large values for this coupling even if $\lam_{345}$ is highly constrained from direct detection and is required to remain relatively small.

\begin{center}
  \begin{figure}[htb!]
    \begin{center}
\includegraphics[width=0.65\textwidth]{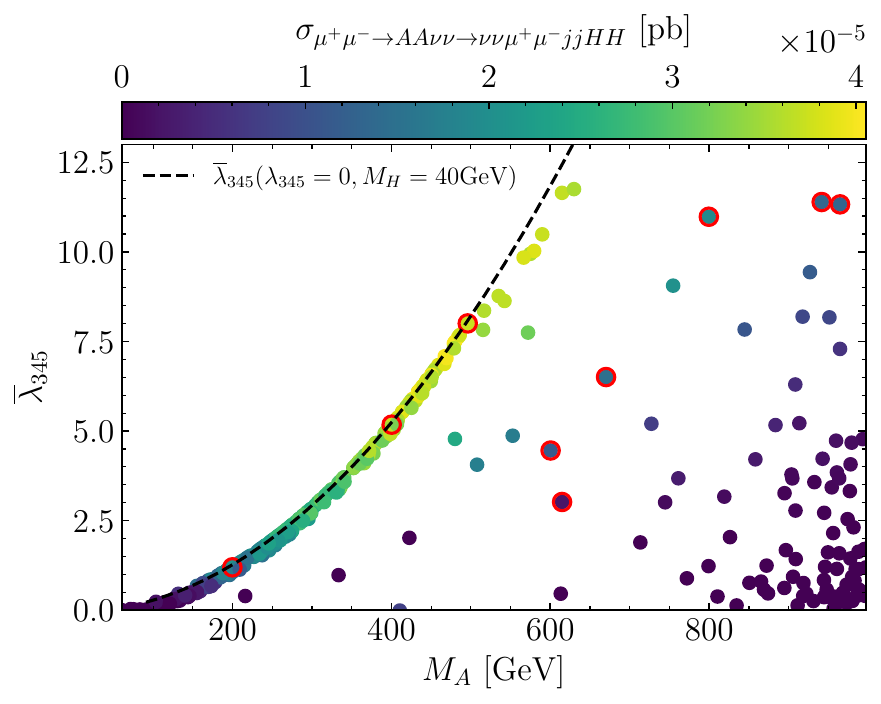}
    \end{center}
    \caption{\label{fig:lam345b} Values for $\bar{\lam}_{345}$ as given by eqn. (\ref{eq:lam345b}) as a function of $M_A$ in a general scan performed in \cite{Braathen:2024ckk}. Also shown are production cross sections for the semileptonic decays of the $A\,A$ system in the VBF type production mode, as well as the upper limit on that coupling for a specific choice of IDM parameters. Investigated benchmark points are shown in red.}
    \end{figure}
\end{center}

Figure \ref{fig:lam345b} gives the values of that coupling in a general scan performed in the above reference. Also shown are cross section values at 10 \TeV~ for the semi-leptonic decay of the $A\,A$ system in the VBF production topology. Investigated benchmark points are given in red.

As before, the full analysis includes a detailed simulation of signal and background, preselection cuts, as well as the training of a neural network used in background suppression. We refer the reader to the original publication for further details. Here we only briefly comment on the results obtained in that work.

\begin{center}
  \begin{figure}[htb!]
    \begin{center}
\includegraphics[width=0.65\textwidth]{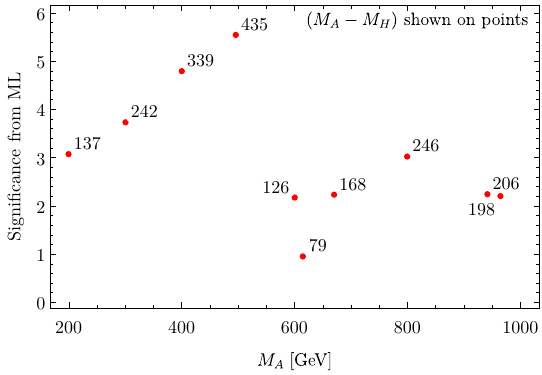}
    \end{center}
    \caption{\label{fig:mamach} Significances obtained using a machine learning algorithm for background suppression, as a function of $M_A$, for selected points. Also shown is the mass difference between $A$ and the dark matter candidate. We see that significances can range up to 6 for specific benchmark points.}
    \end{figure}
\end{center}

\begin{center}
  \begin{figure}[htb!]
    \begin{center}
\includegraphics[width=.49\textwidth]{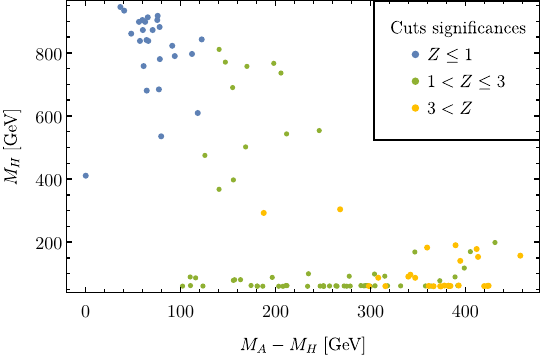}
    \includegraphics[width=.49\textwidth]{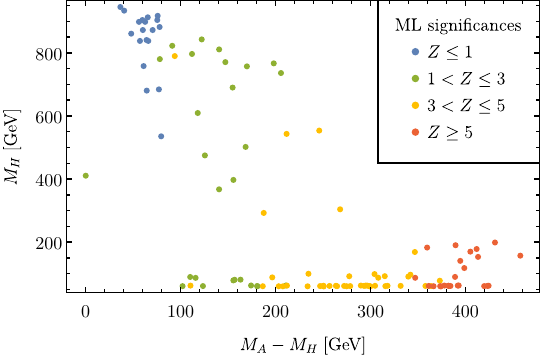}
    \end{center}
    \caption{\label{fig:allres} Significances for a wider range of scan points for the cut based {\sl (left)} as well as machine learning {\sl (right)} analysis, and the $\lb M_A-M_H;\;M_H \rb$ plane. The ML learning result generally supersedes the one based on cuts. Best results are obtained for large mass gaps and low dark matter masses.}
    \end{figure}
\end{center}

In figure \ref{fig:mamach}, we show the significances reached for specific benchmark points as a function of $M_A$ using a machine learning (ML) algorithm. We see that significances up to 6 can be obtained for specific scenarios. The mass gap beteen $A$ and the dark matter candidate is also given.
In figure \ref{fig:allres}, we show the result for a larger spread of points that were generated in the course of the above study. In our approach, the ML method supersedes in general the cut based analysis. Best results were obtained for large mass gaps and low dark matter masses.

\section{Conclusions}
In this proceeding contribution, I briefly discussed new physics scenarios with additional scalar states, in particular in the context of lepton colliders. I first have a short overview on model-independent studies at Higgs factories with center of mass energies of about $250\,\GeV$, and then focussed on a specific model and its discovery prospects at future lepton colliders. The studies and results presented here can only be viewed as examples of new physics scenarios studied at such machines and should serve as a motivation for the experimental collaborations to adept specific searches once the respective experiments have been realized.

\section*{Acknowledgements}
TR acknowledges financial support from the Croatian Science Foundation (HRZZ) project "Beyond the Standard Model discovery and Standard Model precision at LHC Run III", IP-2022-10-2520. This contribution was supported by the National Science Centre, Poland, under the OPUS research project no. 2021/43/B/ST2/01778.
%\begin{thebibliography}{99}
%\bibitem{...}
%....

%\end{thebibliography}
\bibliography{lit,lit_old,lit_lcws}

\end{document}